\documentclass[aps,prl,reprint,showpacs,superscriptaddress]{revtex4-1}

\usepackage{amsmath}
\usepackage{xspace} 
\usepackage{graphicx}

\usepackage{hyperref}
\hypersetup{
  colorlinks   = true, 
  urlcolor     = blue, 
  linkcolor    = black,
  citecolor   = black 
}

\newcommand{\f}[2]{#1 [#2]}  

\newcommand{\ket}[1]{|{#1}\rangle}
\newcommand{\bra}[1]{\langle{#1}|}

\newcommand{\mum}{\,\ensuremath{\mu{\rm m}}\xspace} 
\newcommand{\nm}{\,\ensuremath{{\rm nm}}\xspace}

\newcommand{\MHz}{\,\ensuremath{{\rm MHz}}\xspace}

\newcommand{\ms}{\,\ensuremath{{\rm ms}}\xspace}
\newcommand{\s}{\,\ensuremath{{\rm s}}\xspace}

\makeatletter 
\newcommand\longleftrightarrowfill@{%
  \arrowfill@\leftarrow\relbar\rightarrow}
\newcommand\xleftrightarrow[2][]{%
  \ext@arrow 9999{\longleftrightarrowfill@}{#1}{#2}}
\makeatother

\begin{document}

\title[]{Cavity-Modified Collective Rayleigh Scattering of Two Atoms}

\author{Ren\'{e}~\surname{Reimann}}\email{reimann@iap.uni-bonn.de}
\author{Wolfgang~\surname{Alt}}
\affiliation{Institut f{\"u}r Angewandte Physik, Universit{\"a}t Bonn, Wegelerstra{\ss}e 8, 53115~Bonn, Germany}
\author{Tobias~\surname{Kampschulte}}
\affiliation{Departement Physik, Universit{\"a}t Basel, Klingelbergstrasse 82, 4056~Basel, Switzerland}
\author{Tobias~\surname{Macha}}
\author{Lothar~\surname{Ratschbacher}}
\author{Natalie~\surname{Thau}}
\author{Seokchan~\surname{Yoon}}
\author{Dieter~\surname{Meschede}}
\affiliation{Institut f{\"u}r Angewandte Physik, Universit{\"a}t Bonn, Wegelerstra{\ss}e 8, 53115~Bonn, Germany}


\begin{abstract}
We report on the observation of cooperative radiation of exactly two neutral atoms strongly coupled to the single mode field of an optical cavity, which is close to the lossless-cavity limit. Monitoring the cavity output power, we observe constructive and destructive interference of collective Rayleigh scattering for certain relative distances between the two atoms. Because of cavity backaction onto the atoms, the cavity output power for the constructive two-atom case ($N=2$) is almost equal to the single-emitter case ($N=1$), which is in contrast to free-space where one would expect an $N^2$ scaling of the power. These effects are quantitatively explained by a classical model as well as by a quantum mechanical model based on Dicke states. We extract information on the relative phases of the light fields at the atom positions and employ advanced cooling to reduce the jump rate between the constructive and destructive atom configurations. Thereby we improve the control over the system to a level where the implementation of two-atom entanglement schemes involving optical cavities becomes realistic.
\end{abstract}

\maketitle
Efficient matter-light coupling is a prerequisite for the realization of photonic quantum memories~\cite{Wilk2007, Northup2014} and long-distance quantum networks~\cite{Cirac1997, Kimble2008, Ritter2012}. 
Two paths lead to success for atomic systems:
First, cooperative atom-light interaction, studied more than 50 years ago~\cite{Dicke1953}, has found important applications in quantum information for efficient atom-light interfaces in recent years~\cite{Matsukevich2004, Chou2005, Zhao2008}.
Second, it has become a standard to increase the atom-light interaction by means of cavity quantum electrodynamics (CQED)~\cite{Raimond2006}.
The combination of the two routes by coupling atomic ensembles to cavities recently opened up a rich field of physics:
Cooling~\cite{Ritsch2013}, self-organization of the spatial~\cite{Domokos2002, Black2003, Baumann2010, Arnold2012} and spin degree of freedom~\cite{Dimer2007, Baden2014a}, as well as super- and subradiant phenomena~\cite{Slama2007, Bux2011, Bohnet2012, Keßler2014} have been proposed and observed.

When $N$ free-space atoms cooperatively radiate at higher (smaller) optical power into a certain mode than $N$ independent atoms, the term superradiance (subradiance)~\cite{Gross1982a} is very widely used in literature. 
Here, one should distinguish two cases: 
First, resonant excitation where a significant fraction of the atomic population is excited~\cite{DeVoe1996}. Second,  weak off-resonant Rayleigh scattering with a negligible excited state population~\cite{Inouye1999}, which we consider here. 
In both cases the maximally emitted power scales with $N^2$ for large $N$, which therefore can be seen as a criterion for superradiance.
However, when driving $N$ atoms in a cavity, the atomic dipoles radiate cooperatively into the cavity mode. Simultaneously, the cavity field acts back onto the atoms, which changes their collective nonlinear interaction. In the lossless-cavity limit, the maximally scattered power no longer depends on $N$. 

\begin{figure}
\centering{\includegraphics[width=0.35\textwidth]{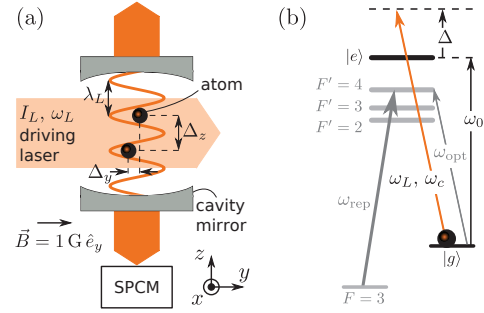}}\\
\caption{(a) Simplified experimental setup. A laser drives two trapped atoms (traps not shown) inside the cavity. The sine curve depicts the atom-cavity coupling strength along the $z$ axis. A single photon counting module (SPCM) detects the photons leaking through the lower cavity mirror. For two atoms inside the cavity, the measured count rate depends on the relative spacings $\Delta_y$ and $\Delta_z$. (b) Atomic level scheme of the $^{133}\text{Cs}$ $\text{D}_2$ line with relevant transitions ($\ket{g}\equiv \ket{F=4, m_F =-4}$ and $\ket{e}\equiv \ket{F'=5, m_F =-5}$). All laser fields are $\sigma^-$ polarized. The cavity mode is $(\sigma^+ + \sigma^-)/\sqrt{2}$ polarized.}
\label{Fig.setup}
\end{figure}

With exactly two neutral atoms strongly coupled to a cavity field our experiment realizes the most elementary and textbooklike situation where both cooperative radiation and cavity backaction become relevant. We show that our measurements are sensitive to the relative atom-field coupling phases and demonstrate how advanced cooling techniques~\cite{Boozer2006, Reiserer2013, Reimann2014a} improve the control over the collective atom-cavity coupling. Now stable relative atom-field coupling phases are obtained for extended periods of time enabling the realization of phase-sensitive cavity-based entanglement schemes for two atoms~\cite{You2003, Metz2006, Kastoryano2011}, which have so far eluded experimental realization.

In our experiment two neutral cesium atoms are captured from background gas by a high-gradient magneto-optical trap (MOT) and are loaded into a red-detuned standing-wave trap at $\lambda_{\rm rDT} = 1030 \nm$. By using the red dipole trap as a conveyor belt the atoms are transported into an orthogonal standing wave, which is formed by the blue-detuned locking light ($\lambda_{\rm bDT}=845.5\nm$) of an optical high-finesse Fabry-P\'{e}rot cavity, for details see~\cite{Reimann2014a}. 
The distances $\Delta_y$ and $\Delta_z$ shown in Fig.~\ref{Fig.setup}(a) are multiples of the lattice periodicities $\lambda_{\rm rDT}/2$ and $\lambda_{\rm bDT}/2$. These atomic distances lead to spatial phase differences along the driving laser and the cavity axes, given by $\phi_{y,z}=2\pi \Delta_{y,z}/\lambda_{L}$, where $\lambda_{L} = 852.3\nm$ is the driving laser wavelength. The combined trapping potential confines the atoms close to the antinodes of the intracavity field. Therefore, the atoms can only realize a $\lambda$ or a $\lambda/2$ pattern~\cite{Fernandez-Vidal2007} along the $z$ axis corresponding to $\phi_z = 0$ or $\phi_z = \pi$, respectively. 
Our CQED system is characterized by the atom-field coupling rate, the cavity field and the atomic population relaxation rates $\{g_0,\, \kappa,\, \Gamma\} \approx 2\pi \times \{18,\, 0.4,\, 5.2    \}\MHz$. The maximum atom-field coupling $g_0$ is calculated from the strongest ($\sigma^-$) dipole transition on the cesium $\rm D_2$ line (for details  see~\cite{Khudaverdyan2008}).

During the measurement three laser fields copropagating along the $y$ axis continuously address the atoms (see Fig.~\ref{Fig.setup}(b)): A strong repumping laser and a weak optical pumping laser (frequencies $\omega_{\rm rep}$ and $\omega_{\rm opt}$) keep most of the atomic population in  the state $\ket{F=4, m_F = -4}$. The driving laser (frequency $\omega_{L} = 2 \pi c /\lambda_{L}$, intensity $I_{L}\approx 2\, {\rm mW}/{\rm cm}^2$, detuning from the atomic resonance $\Delta = \omega_{L} - \omega_0 \approx 2\pi \times 100 \MHz$) is resonant with the cavity ($\delta = \omega_{L} - \omega_c$ = 0). Because of its large detuning, the laser light scatters off the atoms into the cavity mode predominantly by Rayleigh scattering. The cavity output is detected by a single photon counting module (SPCM). 
\begin{figure}
\centering{\includegraphics[width=0.47\textwidth]{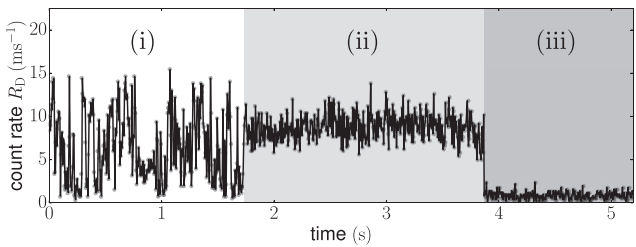}}\\
\caption{Photon count rate of driven atoms coupled to the cavity (single measurement trace). In regions~(i), (ii), (iii) two, one and zero atoms are inside the cavity. The bin time is $5\ms$.}
\label{Fig.2to1to0_atom_trace}
\end{figure}

Fig.~\ref{Fig.2to1to0_atom_trace} shows a SPCM photon count trace, selected to demonstrate the difference between two, one, and zero atoms. In region~(i), where two driven atoms couple to the cavity, the photon count rate jumps between a high ($\approx 12\,{\rm ms}^{-1}$) and a low ($\approx 2\,{\rm ms}^{-1}$) value. The high (low) photon count rates are interpreted as the atoms arranging in a pattern along the $z$ axis that makes the scattered photons interfere constructively (destructively), leading to high collective (low subradiant) emission into the cavity mode. At $1.7\s$ one of the two atoms is lost from the combined trapping potential, leading to a photon count rate with little variance and a mean of about $9\,{\rm ms}^{-1}$ in region (ii). Naively one would expect that two atoms, which interfere constructively, scatter superradiantly and therefore emit 4 times as many photons as a single atom. This is clearly not the case and explained below by the cavity backaction. From $3.9\s$ on (region (iii)), the cavity is empty and only background counts are measured.
 
It is pointed out in~\cite{Tanji-Suzuki2011a} and references therein that many effects in CQED can be fully explained classically when staying in the weak atomic excitation limit. We follow~\cite{Tanji-Suzuki2011a} and  describe the atoms by polarizable particles that radiate as dipoles into a nonquantized cavity field mode: A driving beam with intensity $I_{L}$, which is resonant with the cavity ($\delta = 0$), transversally irradiates $N$ atoms with equal $y$ positions and thus equal driving laser phases. The atoms are assumed to be located within the maxima of the intracavity field with nearest-neighbor distances equal to integer multiples of $\lambda_{L}$ ($\lambda$ pattern). In this situation high collective emission of the atoms into the cavity is expected~\cite{Baumann2010}. In steady state, the field inside the cavity $E_c$ has to reproduce itself after one round trip:
\begin{equation}
\label{Eq.cavity_ss}
E_c = 2 E_\mathcal{M} +r^2E_c.
\end{equation}
Within a round trip the atoms scatter bidirectionally into the cavity mode $\mathcal{M}$ with field amplitude $E_\mathcal{M}$, and the cavity field is reflected from the two mirrors with field reflectivity $r$. The radiation field scattered by the atoms into the cavity mode is described by
\begin{equation}
\label{Eq.E_M}
E_\mathcal{M} = [i N \mathcal{L}(\Delta)/2] ( g_0 \Gamma^{-1}  E_{L} +  g \Gamma^{-1}  2 E_c )  g \tau.
\end{equation}
This field shows contributions from the driving field $E_{L}$ and from the intracavity standing wave $2E_c$. The phase factor $i$ accounts for the dipole emission phase of the $N$ atoms which contribute with their atomic line function $\mathcal{L}(\Delta)=  (-2\Delta \Gamma+ i \Gamma^2) / (\Gamma^2+4\Delta^2)$. The spatially homogeneous driving laser field polarizes the atoms addressing the strongest dipole transition and is therefore weighted with the maximal coupling rate $g_0$. The polarization of the atoms induced by the cavity field scales with the coupling rate $g<g_0$, which is reduced due to the mixed polarization of the cavity mode (see caption Fig.~\ref{Fig.setup}) and due to the fact that the atoms do not necessarily sit in the center of the mode. Both terms are scaled with the atomic lifetime $\Gamma^{-1}$. The last factor in Eq.~(\ref{Eq.E_M}) is interpreted as the polarized atoms acting back onto the cavity field with the coupling rate $g$ within one round trip time  $\tau = 2\ell_0 /c$, where $\ell_0 = 155\mum$ equals the cavity length. 

We insert Eq.~(\ref{Eq.E_M}) into Eq.~(\ref{Eq.cavity_ss}) and solve for the cavity field
\begin{equation}
\label{Eq.E_c_simple}
E_c = -\frac{E_{L}}{2} \frac{g_0}{g} \frac{N}{\frac{i}{2C \mathcal{L}(\Delta)} + N},
\end{equation}
\begin{figure}[!t]
\centering{\includegraphics[width=0.46\textwidth]{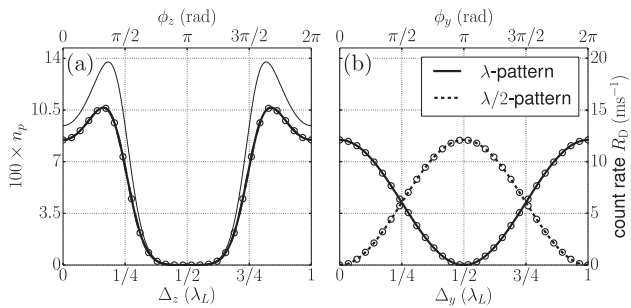}}\\
\caption{Classical (lines) and quantum mechanical (circles) calculations for two driven atoms coupled to a cavity. (a) For $\phi_y = 0$ we show the intracavity photon number $n_p$ for our system parameters (see text, $g= 8 \MHz$, thick line) and, as a comparison, for a lossless cavity ($\kappa = 0$, thin solid line) and an effective free-space situation ($\kappa = 100\,\Gamma$, thin dashed line, in this case the shown $n_p$ is multiplied by $10^{5}$ for better visibility). The free-space scenario illustrates the $N^2$ scaling, see text. The right axis shows the SPCM photon count rate $R_{\rm D}$ for our detection efficiency. (b) Expected $n_p$ and $R_{\rm D}$ as a function of the relative driving laser phase $\phi_y$ for $\lambda$ and $\lambda/2$ patterns ($\phi_z=0$ and $\phi_z=\pi$).}
\label{Fig.Counts_Delta_y_z}
\end{figure}
where we have used the cooperativity $C = g^2/(\kappa \Gamma)$ and  $\kappa = (1-r^2)/\tau$ for $r\approx 1$. The detected SPCM photon count rate is given by 
\begin{equation}
\label{Eq.R_D}
R_{\rm D} = \eta \kappa n_p  =  \eta \kappa \frac{2 \epsilon_0 |E_c|^2 V}{\hbar \omega_{L}},
\end{equation}
where $V = \pi w_c^2 \ell_0 /4$ is the cavity mode volume with the cavity waist $w_c=23\mum$ and $\eta \approx 6\%$ is our overall detection efficiency. We now discuss two limiting cases of Eq.~(\ref{Eq.E_c_simple}). First, in the free-space limit ($\kappa\rightarrow\infty$ and $C\rightarrow 0$) $R_{\rm D}$ scales with $N^2$ as reported for superradiant scattering from a Bose-Einstein condensate in free space~\cite{Inouye1999}. Other experiments performed with ions in free space~\cite{Eichmann1993} or ions interacting with their mirror images~\cite{Eschner2001} also show a strong $N$ dependence of the detected signal. However, in the lossless-cavity limit ($\kappa\rightarrow 0$ and $C\rightarrow \infty$) this dependency becomes negligible due to the cavity backaction: The intracavity field builds up $\pi$ shifted with respect to the phase of the driving field such that the two fields completely cancel at the positions of the $N$ atoms, thus reducing the $N^2$ scaling to an $N$-independent scattering rate. Cavity backaction is described in detail in~\cite{Alsing1992} for one atom in a lossless cavity. Our system is close to this limit, thereby showing only a small difference in detected photon counts between the one- and the two-atom case (see Fig.~\ref{Fig.2to1to0_atom_trace} and Table~\ref{tab.mini}).
\begin{table}[h]
\begin{center}
\begin{tabular}{c c c c c c c c c c}
\hline
\hline
\rule{0pt}{2.5ex}  Count rate $(\rm ms^{-1})$& & & Measurement & & &   Model   & & &  Free space \\
\hline
One atom & & &  $9(1)$  & & &  $9.5$  & & &  $9$ \\
Two atoms & & &  $12(2)$  & & &  $12.1$ & & &  $36$\\
\hline
\hline
\end{tabular}
\end{center}
\caption{Measured and calculated count rates for one and two atoms. We compare the data from Fig.~\ref{Fig.2to1to0_atom_trace} to the classical model for two constructively interfering atoms with $g= 8 \MHz$ as a free parameter. The free space scenario predicts a fourfold two-atom signal. Here, we assume the experimental one-atom photon count rate.}
\label{tab.mini} 
\end{table}

To understand our measurement signals, we consider the general, position-dependent model with arbitrary atomic positions along the $y$ and the $z$ axis and a finite laser-cavity detuning $\delta$ and write down the intracavity field~\cite{Tanji-Suzuki2011a}
\begin{equation}
\label{Eq.E_c_full}
E_c = -\frac{E_{L}}{2} \frac{g_0}{g} \frac{N \mathcal{G}}{   \frac{i}{2 C \mathcal{L}(\Delta)} + \frac{\delta}{2 \kappa C \mathcal{L}(\Delta)} + N \mathcal{H}}.
\end{equation}
\begin{figure}
\centering{\includegraphics[width=0.42\textwidth]{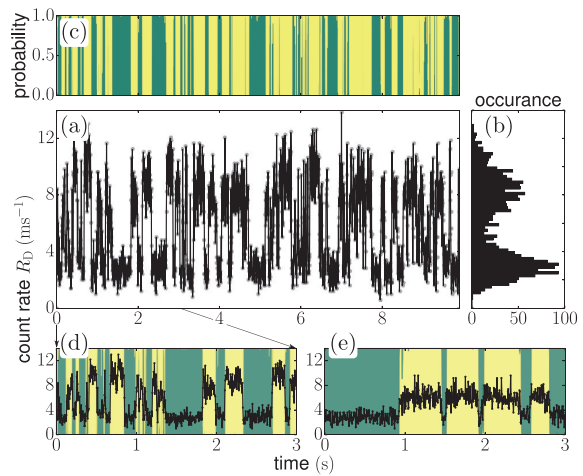}}\\
\caption{Motional dynamics between collective states of two atoms. (a),(b) The bimodal structure of the count rate data arising from quantized atom hopping along the $z$ direction is clearly visible in the time trace and in the corresponding histogram. (c) HMM probabilities (see text) corresponding to trace (a) for the constructively and destructively emitting two atom states in yellow and green, respectively. (d) and (e) compare data and HMM probabilities (background) with cavity cooling only, and with additional continuous Raman sideband cooling, respectively.}
\label{Fig.HMM_time_trace_data}
\end{figure}
The collective coupling parameter for the cavity is given by $\mathcal{H} = \frac{1}{N}\sum_{n=1}^{N} \f{\cos^2}{(2\pi /\lambda_{L}) z_n}$, while the collective coupling parameter for the driving beam is described by $\mathcal{G} = \frac{1}{N}\sum_{n=1}^{N} \f{\exp}{i (2\pi /\lambda_{L}) y_n} \f{\cos}{(2\pi /\lambda_{L}) z_n}$. For $N=2$ atoms, one of which is positioned at maximum cavity coupling, these parameters take the form $\mathcal{H} = \frac{1}{2}[1+ \cos^2(\phi_z)]$ and  $\mathcal{G} = \frac{1}{2}[1+ \exp(i \phi_y) \cos(\phi_z)]$. 
Based on Eqs.~\eqref{Eq.R_D} and \eqref{Eq.E_c_full} we now analyze the different mean values of the high and low count rate levels in our data. Thereby we find---in good agreement with~\cite{Reick2010}---that the atoms couple with an effective $g$ between $8$ and $10\MHz$, depending on the radial atom positions within the cavity mode in each experimental repetition. 

Curves calculated from Eqs.~\eqref{Eq.R_D} and \eqref{Eq.E_c_full} are shown in Fig.~\ref{Fig.Counts_Delta_y_z}. The values of $n_p$ and $R_{\rm D}$ for two atoms with $\phi_y=0$ are depicted in Fig.~\ref{Fig.Counts_Delta_y_z}(a). One atom is fixed at an antinode of the cavity field. As the other atom is moved along the intracavity standing wave, the photon count rate first rises, compared to $\Delta_z = 0$. At $\Delta_z = \lambda_{L}/2$ no cavity output is expected due to destructive interference between the two emitters. In Fig.~\ref{Fig.Counts_Delta_y_z}(a) the values for a single atom are described by the curves at position $\Delta_z = \lambda_{L}/4$. Comparing $n_p$ at $\Delta_z = 0$ (constructive interference for two cavity-coupled atoms) to $\Delta_z = \lambda_{L}/4$ (one cavity-coupled atom) shows that our system is close to the lossless-cavity limit with strong cavity backaction. In contrast, we show the expected $n_p$ for an open cavity close to the free-space limit. Here superradiant scattering at $\Delta_z = 0$ with four times the single-atom emission is calculated, recovering the $N^2$ scaling. 

Figure~\ref{Fig.Counts_Delta_y_z}(b) explains our measured two-atom data (Figs.~\ref{Fig.2to1to0_atom_trace}(i) and \ref{Fig.HMM_time_trace_data}). For each single trace two atoms are loaded randomly into our conveyor belt, realizing an arbitrary but fixed relative distance $\Delta_y$ and phase $\phi_y$ along the $y$ direction. The measured data show that the atoms jump back and forth between the $\lambda$ and the $\lambda/2$ pattern ($\phi_z=0$ and $\phi_z=\pi$), while $\Delta_y$ and $\phi_y$ remain fixed during a single trace, as indicated by the constant upper and lower count rate levels. Besides traces such as Fig.~\ref{Fig.2to1to0_atom_trace}(i), where $\phi_y$ is close to 0 or $\pi$ (maximal jump contrast), we do also observe shots with lower (see Fig.~\ref{Fig.HMM_time_trace_data}) or even vanishing jump contrast. The latter correspond to $\phi_y$ near $\frac{\pi}{2}$ or $\frac{3\pi}{2}$. 

The observation of hopping along the $z$ but not along the $y$ direction is explained by the fact that the dominating heating mechanism, parametric heating due to amplitude fluctuations of the intracavity trapping field, is strong along the $z$ axis but negligible along the $y$ axis~\cite{Gehm1998, Reimann2014a}. This claim is supported by the fact that 1D Raman cooling along the $z$ axis significantly reduces the hopping rate; see Fig.~\ref{Fig.HMM_time_trace_data}. To extract the jump rates from our noisy signals, we apply to our data a hidden Markov model (HMM) approach~\cite{Gammelmark2014} with a time resolution of $50\,{\rm \mu s}$, which is much faster than the inverse jump rates. The algorithm is based on two hidden states and calculates the probabilities of being in the constructive and destructive emission pattern. While Fig.~\ref{Fig.HMM_time_trace_data}(d) is measured under standard cavity cooling conditions~\cite{Zippilli2005a}, Fig.~\ref{Fig.HMM_time_trace_data}(e) shows data where an additional Raman sideband cooling beam is continuously irradiating the atoms. By cooling on the motional $z$ sideband~\cite{Reimann2014a}, we counteract the parametric heating and reduce the jump rate by a factor $\gtrsim 5$. For both cooling scenarios the jump rates from the constructively to the destructively emitting state and vice versa are equal within the experimental uncertainty. This indicates that the dynamics are governed by thermal excitations and not by collective forces which can lead to self-ordering~\cite{Domokos2002}.
  
Our experimental situation is characterized by a negligible population ($<10^{-3}$) in the excited atomic states $\ket{e}_n$ for all atoms $n$ and therefore well described by the classical theory. However, a quantum mechanical model can reveal the involved joint atomic states and their symmetries. Here, the Tavis-Cummings Hamiltonian of two driven atoms  inside the cavity is given by
\begin{equation}
\hat{H} = \hat{H}_{\rm at} + \hat{H}_{\rm cav} +\hat{H}_\text{at-cav} +\hat{H}_{L},
\label{Eq.H}
\end{equation}
with $\hat{H}_{\rm at} = -\hbar \Delta \sum_{n=1,2} \hat{\sigma}_n^\dagger \hat{\sigma}_n$ and $\hat{H}_{\rm cav} = -\hbar \delta \, \hat{a}^\dagger \hat{a}$ being the atomic and cavity Hamiltonian, respectively. The atomic lowering operators are described by $\hat{\sigma}_n = \ket{g}_n \bra{e}$ and the cavity field operator annihilating one photon is $\hat{a}$.  For direct comparison between the quantum mechanical and the classical approach we numerically solve~\cite{Johansson2013} the system's master equation~\cite{Zippilli2004} and display the results as circles in Fig.~\ref{Fig.Counts_Delta_y_z}. 

We use the system's symmetry~\cite{Fernandez-Vidal2007} to write the atom-cavity term and the drive term in Eq.~(\ref{Eq.H}) as $\hat{H}_\text{at-cav}  = \hat{H}_+ + \hat{H}_-$ with $\hat{H}_\pm = \hbar g_\pm (\hat{a}\hat{S}_\pm^\dagger + \hat{a}^\dagger\hat{S}_\pm)$, $g_\pm = g[1 \pm \cos(\phi_z)]/\sqrt{2}$ and $\hat{H}_{L} = (\hbar \sqrt{2} \Omega_{L}/2)[ \cos(\phi_y / 2) \hat{S}_+ + i \sin(\phi_y / 2) \hat{S}_-  + \text{H.c.} ]$ with the Rabi frequency $\Omega_{L} = \Gamma \sqrt{I_L/(2I_{\rm sat})}$ and $I_{\rm sat}\approx 1.1 \, {\rm mW}/{\rm cm}^2$. The Hermitian conjugates of the Dicke operators $\hat{S}_\pm = (\hat{\sigma}_1 \pm \hat{\sigma}_2)/\sqrt{2}$ create the symmetric and antisymmetric Dicke states according to $\hat{S}_\pm^\dagger \ket{gg} = \ket{\pm} = (\ket{eg} \pm \ket{ge})/\sqrt{2}$. 
\begin{figure}
\includegraphics[width=0.475\textwidth]{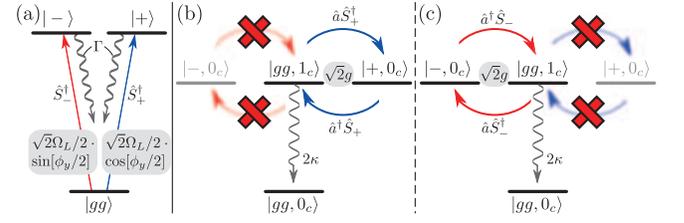}
\caption{Quantum mechanical picture of collective scattering of two atoms in a cavity. (a)~Driving laser pumping the Dicke states. As a function of the relative driving phase $\phi_y$ the symmetric $\ket{+}$ or antisymmetric $\ket{-}$ Dicke state is weakly excited. (b)~Coupling to the cavity for $\lambda$ pattern, $\phi_z = 0$. (c)~Coupling to the cavity for $\lambda/2$ pattern, $\phi_z = \pi$. Depending on the atomic pattern, only $\ket{+}$ or only $\ket{-}$ couples to the cavity. If the cavity-coupled Dicke state is pumped by the driving laser, the cavity mode is populated with photons ($\ket{0_c} \rightarrow \ket{1_c}$), which leave the system via the cavity loss channel $2\kappa$ and can be detected by the SPCM.}
\label{Dicke_Levels}
\end{figure}

The advantage of writing the interaction terms of the Hamiltonian in this form becomes clear with Fig.~\ref{Dicke_Levels}, which pictures the dynamics driven by $\hat{H}_{L}$ and $\hat{H}_\text{at-cav}$. Here, the cavity backaction can be explained by the quantum-path interference \mbox{$\ket{gg,0_c} \xleftrightarrow{\hat{H}_{L}} \ket{\pm,0_c} \xleftrightarrow{\hat{H}_\text{at-cav}} \ket{gg,1_c} $}, which leads to a suppression of atomic excitation~\cite{Fernandez-Vidal2007} and therefore to less scattering than in the free-space limit. We note that the collective coupling and driving strengths are increased by a factor of $\sqrt{2}$ compared to the single atom frequencies $g$ and $\Omega_{L} /2$. This $\sqrt{N}$ enhancement is typical for driving Dicke states $\ket{\pm}$~\cite{Dicke1953, Dimer2007} and leads to the $N^2$ scaling of the scattered power in the free-space limit.

In conclusion, we have demonstrated the most elementary case of cooperative coupling of atoms to an optical cavity. Operating close to the lossless-cavity limit, the backaction of the cavity field onto the atoms leads to a strong modification of the constructive emission: Two atoms scatter about the same intensity into the cavity mode as a single emitter. Additional Raman cooling leads to stable coupling and constant phases ($\phi_y$, $\phi_z$) on the few hundred ms scale, which is long compared to typical coherence times and gate-operation times of neutral atom cavity experiments. The demonstrated two-atom control paves the way to phase sensitive entanglement schemes for two neutral atoms coupled to an optical cavity~\cite{You2003, Metz2006, Kastoryano2011}. 

After submission of this manuscript we have learned of related recent work with two ions coupled to a cavity~\cite{Casabone2015}.

\begin{acknowledgments}
We wish to thank S. Gammelmark for providing his HMM code. We acknowledge financial support by the Q.com-Q Verbundprojekt of the BMBF and by the Integrated Project SIQS of the European Commission. L.\,R. acknowledges support from the Alexander von Humboldt Foundation. N.\,T. and R.\,R. acknowledge support by the Bonn-Cologne Graduate School of Physics and Astronomy. 
\end{acknowledgments}
%
%

%
\end{document}